\documentclass[12pt]{article}

\begin{document}

\title{\bf Addendum: A Classification of Plane Symmetric Kinematic
Self-similar Solutions}

\author{M. Sharif \thanks{msharif@math.pu.edu.pk} and Sehar Aziz
\thanks{sehar$\_$aziz@yahoo.com}\\
Department of Mathematics, University of the Punjab,\\
Quaid-e-Azam Campus, Lahore-54590, Pakistan.}

\date{}

\maketitle
In our recent paper, we classified plane symmetric kinematic
self-similar perfect fluid and dust solutions of the second, zeroth
and infinite kinds. However, we have missed some solutions during
the process. In this short communication, we add up those missing
solutions. We have found a total of seven solutions, out of which
five turn out to be independent and cannot be found in the earlier
paper.

Recently, we presented a classification of kinematic self-similar
plane symmetric spacetimes \cite{jkps2}. We have discussed the plane
symmetric solutions that admit kinematic self-similar vectors of the
second, zeroth, and infinite kinds when the perfect fluid is tilted
to the fluid flow, parallel or orthogonal. However, we missed some
cases that could provide more solutions. In this addendum, we
present those missing solutions, which turn out to be five in
number. Further, for the the self-similarity of the first kind
(tilted), the two-fluid formalism does not work as the self-similar
variable is $\xi=\frac{x}{t}$. We shall investigate a different
approach to obtain the solution in this case. The tilted perfect
fluid yields four more solutions (one first-kind solution, two
2nd-kind solutions and one zeroth-kind solution), the parallel
perfect fluid gives one infinite kind solution, and the orthogonal
perfect fluid provides two solutions (one first-kind solution and
one 2nd-kind solution). Thus, we obtain total seven solutions out of
which five solutions are independent. The plane symmetric metric
considered in the paper \cite{jkps2} is the following:
\begin{equation}\label{E:P1}
ds^2=e^{2\nu(t,x)}dt^2-dx^2-e^{2\lambda(t,x)}(dy^2+ dz^2).
\end{equation}
We are skipping the details as the procedure can be seen elsewhere
\cite{jkps2}.

The tilted perfect fluid of the first kind implies that the energy
density $\rho$ and pressure $p$ must take the following forms:
\begin{eqnarray}\label{E:PdSST1}
\kappa\rho &=& \frac{1}{x^2}\rho(\xi),\\\label{E:PpSST1} \kappa p &=&
\frac{1}{x^2}p(\xi),
\end{eqnarray}
where the self-similar variable is $\xi=x/t$. When the Einstein
field equations (EFEs) and the equations of motion for the matter
field are satisfied, a set of ordinary differential equations (ODEs)
is obtained, hence, the EFEs and equations of motion~\cite{jkps2}
reduce to
\begin{eqnarray}\label{E:PSElnT11}
\dot{\rho} &=&-2\dot{\lambda}(\rho+p),\\\label{E:PSElnT12} 2p-\dot{p} &=&
\dot{\nu}(\rho+p),\\\label{E:PSElnT13} \rho &=& -4\dot{\lambda}
-3{\dot{\lambda}}^{2}-2\ddot{\lambda}-1,\\\label{E:PSElnT14} 0&=&
{\dot{\lambda}}^{2},\\\label{E:PSElnT15}
0&=&\ddot{\lambda}+{\dot{\lambda}}^2+\dot{\lambda}
-\dot{\lambda}\dot{\nu},\\\label{E:PSElnT16} p&=&1+2\dot{\lambda}+{\dot{\lambda}}^2
+2\dot{\nu}+2\dot{\lambda}\dot{\nu},\\\label{E:PSElnT17} 0&=& 2\dot{\lambda}\dot{\nu}
-2\ddot{\lambda}-3{\dot{\lambda}}^2-2\dot{\lambda},\\\label{E:PSElnT18} p &=&
\ddot{\lambda}+{\dot{\lambda}}^2 +\dot{\lambda}
+\dot{\lambda}\dot{\nu}+\ddot{\nu}+{\dot{\nu}}^2,\\\label{E:PSElnT19}
0&=&-\ddot{\lambda}-{\dot{\lambda}}^2 -\dot{\lambda}+\dot{\lambda}\dot{\nu}.
\end{eqnarray}
Only the EOS(3) is compatible with this kind. Equations
(\ref{E:PdSST1}) and (\ref{E:PpSST1}) yield $p=k\rho$ while Eqs.
(\ref{E:PSElnT14}) and (\ref{E:PSElnT11}) imply that $\lambda$ and
$\rho$, respectively, are arbitrary constants. Also, Eq.
(\ref{E:PSElnT12}) gives $\dot{\nu}=\frac{2k}{k+1}$. Using this
value in the remaining equations, we obtain the following solution:
\begin{eqnarray}
\nu&=&\ln{(c_0\xi^{(1\mp\sqrt{2})})},\quad \lambda=c_2,\nonumber\\\label{E:PTPF1S1}
\rho&=&constant,\quad k=-3\pm\sqrt{2}.
\end{eqnarray}
The corresponding metric is
\begin{equation}\label{E:PTPF1S1CM}
ds^2=(\frac{x}{t})^{(2\mp2\sqrt{2})}dt^2-dx^2-x^2(dy^2+ dz^2).
\end{equation}

For the self-similarity of the second kind, we obtain solutions only
with the EOS(3), and these solutions are missing in Ref. 1. The
energy density $\rho$ and pressure $p$ can be written as
\begin{eqnarray}\label{E:PdSST2}
\kappa\rho &=& \frac{1}{x^2}[\rho_1(\xi)+\frac{x^2}{t^2}\rho_2(\xi)],\\
\label{E:PpSST2}\kappa p &=& \frac{1}{x^2}[p_1(\xi)+\frac{x^2}{t^2}p_2(\xi)],
\end{eqnarray}
where the self-similar variable is $\xi=x/(\alpha
t)^\frac{1}{\alpha}$. A set of ODEs yield
\begin{eqnarray}\label{E:PSElnT21}
\dot{\rho_1} &=&-2\dot{\lambda}(\rho_1+p_1),\\\label{E:PSElnT22}
\dot{\rho_2}+2\alpha\rho_2
&=&-2\dot{\lambda}(\rho_2+p_2),\\\label{E:PSElnT23} -\dot{p_1}+2p_1
&=& \dot{\nu}(\rho_1+p_1),\\\label{E:PSElnT24} -\dot{p_2}
&=&\dot{\nu}(\rho_2+p_2),\\\label{E:PSElnT25} \rho_1 &=&
-4\dot{\lambda}
-3{\dot{\lambda}}^{2}-2\ddot{\lambda}-1,\\\label{E:PSElnT26}
\alpha^2e^{2\nu}\rho_2 &=& {\dot{\lambda}}^{2},\\\label{E:PSElnT27}
0&=&\ddot{\lambda}+{\dot{\lambda}}^2+\dot{\lambda}
-\dot{\lambda}\dot{\nu},\\\label{E:PSElnT28}
p_1&=&1+2\dot{\lambda}+{\dot{\lambda}}^2+2\dot{\nu}
+2\dot{\lambda}\dot{\nu},\\\label{E:PSElnT29} \alpha^2e^{2\nu}p_2
&=& -2\ddot{\lambda}-3{\dot{\lambda}}^2
-2\alpha\dot{\lambda}+2\dot{\lambda}\dot{\nu},\\\label{E:PSElnT210}
p_1&=&\ddot{\lambda}+{\dot{\lambda}}^2
+\dot{\lambda}+\dot{\lambda}\dot{\nu}+\ddot{\nu}+{\dot{\nu}}^2,\\\label{E:PSElnT211}
\alpha^2e^{2\nu}p_2&=&-\ddot{\lambda}-{\dot{\lambda}}^2
-\alpha\dot{\lambda}+\dot{\lambda}\dot{\nu}.
\end{eqnarray}
Proceeding along the same lines with the EOS(3), as given in Ref. 1,
for $k\neq-1$ we assume that $\rho_1=0$ and that $\rho_2$ is
arbitrary. Thus, Eqs. (\ref{E:PSElnT25}), (\ref{E:PSElnT27}), and
(\ref{E:PSElnT28}) show that $\dot{\nu}=0$ and $\dot{\lambda}=-1$,
and Eq. (\ref{E:PSElnT22}) implies that $\alpha={k+1}$. Equations
(\ref{E:PSElnT29}) and (\ref{E:PSElnT211}) give $\alpha=2$, and we
obtain the following spacetime:
\begin{eqnarray}
\nu&=&c_1, \quad \lambda=-\ln{\xi}+c_2, \quad \rho_1=0=p_1,\quad
\rho_2=constant=p_2,\nonumber\\\label{E:PTPF2S2S} \alpha&=&2,\quad k=1.
\end{eqnarray}
The resulting plane symmetric metric becomes
\begin{equation}\label{E:PTPF2S2CMS}
ds^2=dt^2-dx^2-2t(dy^2+ dz^2).
\end{equation}
When $k\neq-1$, we take $\rho_2=0$, and $\rho_1$ is arbitrary;
hence, Eq. (\ref{E:PSElnT26}) implies that $\dot{\lambda}=0$. For
the first possibility, it follows that
\begin{eqnarray}
\nu&=&\frac{2k}{k+1}\ln{\xi}+c_1,\quad \lambda=c_2,\quad p_1=k\rho_1
, \quad \rho_1=constant, \nonumber\\\label{E:PTPF2S3}
p_2&=&0=\rho_2, \quad k=-3\pm2\sqrt{2};
\end{eqnarray}
hence, the plane symmetric spacetime will take the following form:
\begin{equation}\label{E:PTPF2S3CM}
ds^2=(\frac{x}{(\alpha t)^{1/\alpha}})^\frac{4k}{k+1}dt^2-dx^2-x^2(dy^2+ dz^2).
\end{equation}

For the self-similarity of the zeroth kind, the EFEs show that the
quantities $\rho$ and $p$ must be of the form
\begin{eqnarray}\label{E:PdSST0}
\kappa\rho &=& \frac{1}{x^2}[\rho_1(\xi)+x^2\rho_2(\xi)],\\
\label{E:PpSST0} \kappa p &=& \frac{1}{x^2}[p_1(\xi)+x^2p_2(\xi)],
\end{eqnarray}
where the self-similar variable is $\xi=\frac{x}{e^{t}}$. A set of
ODEs follows such that
\begin{eqnarray}\label{E:PSElnT01}
\dot{\rho_1} &=&-2\dot{\lambda}(\rho_1+p_1),\\\label{E:PSElnT02} \dot{\rho_2}
&=&-2\dot{\lambda}(\rho_2+p_2),\\\label{E:PSElnT03} -\dot{p_1}+2p_1 &=&
\dot{\nu}(\rho_1+p_1),\\\label{E:PSElnT04} -\dot{p_2}
&=&\dot{\nu}(\rho_2+p_2),\\\label{E:PSElnT05}\rho_1 &=&
-4\dot{\lambda}-3{\dot{\lambda}}^{2}-2\ddot{\lambda}-1,\\\label{E:PSElnT06}
e^{2\nu}\rho_2 &=& {\dot{\lambda}}^{2},\\\label{E:PSElnT07}
0&=&\ddot{\lambda}+{\dot{\lambda}}^2
+\dot{\lambda}-\dot{\lambda}\dot{\nu},\\\label{E:PSElnT08}
p_1&=&1+2\dot{\lambda}+{\dot{\lambda}}^2
+2\dot{\nu}+2\dot{\lambda}\dot{\nu},\\\label{E:PSElnT09}
e^{2\nu}p_2&=&2\dot{\lambda}\dot{\nu}
-2\ddot{\lambda}-3{\dot{\lambda}}^2,\\\label{E:PSElnT010}
p_1&=&\ddot{\lambda}+{\dot{\lambda}}^2 +\dot{\lambda}+\dot{\lambda}\dot{\nu}
+\ddot{\nu}+{\dot{\nu}}^2,\\\label{E:PSElnT011}
e^{2\nu}p_2&=&-\ddot{\lambda}-{\dot{\lambda}}^2 +\dot{\lambda}\dot{\nu}.
\end{eqnarray}
For the EOS(3) when $k\neq-1,~\rho_2=0$, and $\rho_1$ is arbitrary,
Eq. (\ref{E:PSElnT06}) yields $\dot{\lambda}=0$ while Eqs.
(\ref{E:PSElnT01}) and (\ref{E:PSElnT03}) show that
$\dot{\nu}=\frac{2k}{k+1}$. Finally, we obtain the same solution as
in the case of the second kind with the EOS(3) given by Eq.
(\ref{E:PTPF2S3}) with $\alpha=0$. The corresponding metric is
\begin{equation}\label{E:PTPF0S4CM}
ds^2=(xe^{-t})^\frac{4k}{k+1}dt^2-dx^2-e^{2t} (dy^2+ dz^2).
\end{equation}

For the self-similarity of the first kind in the orthogonal perfect
fluid case, the EFEs and the equations of motion give
\begin{eqnarray}\label{E:PSElnO12} e^{2\nu}(1+\rho)&=&
{\lambda'}^2,\\\label{E:PSElnO13}
e^{2\nu}(3-p)&=&3{\lambda'}^2+2\lambda''-2\lambda'\nu',\\\label{E:PSElnO14}
e^{2\nu}(1-p)&=&\lambda''+{\lambda'}^2-\lambda'\nu',\\\label{E:PSElnO15}
2\lambda'(\rho+p)&=&-\rho',\\\label{E:PSElnO16} \rho&=&p.
\end{eqnarray}
Clearly, Eq. (\ref{E:PSElnO16}) shows that this is a system with a
stiff fluid. If these equations are solved simultaneously, Eq.
(\ref{E:PSElnO15}) provides the value of $\lambda'$, and Eq.
(\ref{E:PSElnO12}) gives the value of $\nu$ in terms of $p$.
Equations (\ref{E:PSElnO13}) and (\ref{E:PSElnO14}) impose a
constraint on $p$, ${p'}^2p-2(1+p)(p''p-{p'}^2)=0$, and we arrive at
the following solution:
\begin{eqnarray}
\nu&=&\ln{(\frac{p'}{4p\sqrt{(1+p)}})},\quad
\lambda=-\frac{1}{4}\ln{(p)}+\ln{(c_1)},\quad \rho=p\label{E:POPF1S1}.
\end{eqnarray}

For the self-similarity of the second kind in the orthogonal perfect
fluid case, the quantities $\rho$ and $p$ must be of the forms
\begin{eqnarray}\label{E:POd21}
\kappa\rho
&=&x^{-2}\rho_1(\xi)+x^{-2\alpha}\rho_2(\xi),\\\label{E:POp21}
\kappa p &=&x^{-2}p_1(\xi)+x^{-2\alpha}p_2(\xi).
\end{eqnarray}
A set of ODEs gives
\begin{eqnarray}\label{E:PSElnO21}
\rho'_1&=& -2\lambda'(\rho_1+p_1),\\\label{E:PSElnO22} \rho'_2&=&
-2\lambda'(\rho_2+p_2),\\\label{E:PSElnO23}
2p_1&=&\alpha(\rho_1+p_1),\\\label{E:PSElnO24} \rho_2 &=& p_2\\\label{E:PSElnO25}
\rho_1 &=& -1,\\\label{E:PSElnO26} e^{2\nu}\rho_2 &=&
{\lambda'}^2,\\\label{E:PSElnO27} 0 &=& (1-\alpha)\lambda',\\\label{E:PSElnO28} p_1
&=& 1+2\alpha,\\\label{E:PSElnO29}
e^{2\nu}p_2&=&-2\lambda''+2\lambda'\nu'-3{\lambda'}^2,\\\label{E:PSElnO210}
p_1&=&\alpha^2,\\\label{E:PSElnO211}
e^{2\nu}p_2&=&-\lambda''-{\lambda'}^2+\lambda'\nu'.
\end{eqnarray}
Equation (\ref{E:PSElnO24}) represents a stiff fluid, and Eq.
(\ref{E:PSElnO27}) gives $\lambda'=0$; hence, we obtain the
following solution:
\begin{eqnarray}
\nu&=&arbitrary,\quad \lambda=c_4,\quad p_2=0=\rho_2,\quad
\rho_1=-1,\nonumber\\\label{E:POPF2S1}p_1&=&3\pm2\sqrt{2},\quad \alpha=1\pm\sqrt{2}.
\end{eqnarray}
The corresponding metric is
\begin{equation}\label{E:POPF2S1CM}
ds^2=x^{2(1\pm\sqrt{2})}dt^2-dx^2-x^2(dy^2+dz^2).
\end{equation}

For the self-similarity of the infinite kind in the parallel perfect
fluid, a set of ODEs is given as
\begin{eqnarray} \label{E:PSEPinf1}
-\rho&=& 3{\lambda'}^2+2\lambda'',\\\label{E:PSEPinf2} p
&=&{\lambda'}^2+2\lambda'\nu',\\\label{E:PSEPinf3} p
&=&\lambda''+{\lambda'}^2+\lambda'\nu'+\nu''+{\nu'}^2,\\ \label{E:PSEPinf4}
-p'&=&\nu'(\rho+p).
\end{eqnarray}
Solving Eqs. (\ref{E:PSEPinf1})-(\ref{E:PSEPinf4}) with the
assumption that $p$ is constant, we find that $\lambda$ is a linear
function of $\xi$. Finally, we arrive at the following solution:
\begin{eqnarray}
\nu&=&c_1,\quad \lambda=c_2\xi+c_3,\nonumber\\
\label{E:PPPinfS1}\rho&=&-3p=constant.
\end{eqnarray}
The metric is given by
\begin{equation}\label{E:PPPinfS1CM}
ds^2=dt^2-dx^2-e^{2x}(dy^2+ dz^2).
\end{equation}

We notice that the solutions given by Eqs.(14), (31) and (66) turn
out to be dependent and the solutions given by Eqs. (29), (45),
(51), and (72) are independent. Thus, we have a total of five
independent solutions. It is worth mentioning here that the
self-similar solutions in Eqs. (14), (31), and (66) correspond to
the already classified solutions \cite{jmp} under particular
coordinate transformations. The metrics given by Eqs. (14), (31),
and (66) correspond to the class of metrics
\begin{equation}
ds^2=e^{2\nu(x)}dt^2-dx^2-e^{2\lambda(x)}(dy^2+ dz^2),
\end{equation}
which has four Killing vectors admitting $G_3\otimes\Re$ with a
timelike $\Re$. We also note that the density is either zero or
positive in all the solutions, except for the solution given by Eq.
(66). The physical properties of all these solutions can be seen in
the Ref. 4.

We would like to mention here that the paper in Ref. 1 focussed on a
classification of plane symmetric kinematic self-similar solutions
under certain assumptions. A complete classification for the most
general plane symmetric kinematic self-similar solutions appears
elsewhere \cite{cqg1}.

\vspace{0.5cm}

\begin{description}
\item  {\bf ACKNOWLEDGMENT}
\end{description}

One of us (SA) would like to acknowledge Higher Education Commission
(HEC) for the financial support.

\vspace{0.5cm}


\end{document}